\documentclass[fleqn,usenatbib]{mnras}

\usepackage{newtxtext,newtxmath}

\usepackage[T1]{fontenc}
\usepackage{ae,aecompl}

\usepackage{graphicx}	
\usepackage{amsmath}	
\usepackage{amssymb}	

\title[First TNO in polar resonance with Neptune]{First transneptunian object in polar resonance with Neptune}

\author[]{
M.H.M. Morais$^{1}$\thanks{E-mail: helena.morais@rc.unesp.br (MHMM)}
F. Namouni,$^{2}$\thanks{E-mail: namouni@obs-nice.fr (FN)}
\\
$^{1}$Universidade Estadual Paulista (UNESP), Instituto de Geoci\^encias e Ci\^encias Exatas, Av. 24-A, 1515, 13506-900 Rio Claro, SP, Brazil \\
$^{2}$Universit\'e C\^ote d'Azur, CNRS, Observatoire de la C\^ote d'Azur, CS 24229, 06304 Nice, France
}

\date{Accepted XXX. Received YYY; in original form ZZZ}

\pubyear{2017}

\begin{document}
\label{firstpage}
\pagerange{\pageref{firstpage}--\pageref{lastpage}}
\maketitle

\begin{abstract}

Capture in mean motion resonance has been observed in the Solar System for small objects with prograde as well as retrograde orbits of moderate inclinations. However, no example of an object with a nearly polar orbit was known to be in resonance with a planet. In this Letter, we report   that the nearly-polar transneptunian object (471325), nicknamed  Niku, is in a 7:9 resonance with Neptune, with a mean lifetime in resonance of  $16\pm11$ million years. While entrance and exit in the 7:9 resonance is caused by close encounters with Neptune the resonant configuration provides a temporary protection mechanism against disruptive close encounters with this planet. The other nearly polar transneptunian objects do not seem to be in resonance with the planets with the possible exception of 2008 KV42, also known as Drac, that has a small chance of being in the 8:13 resonance with Neptune.  
\end{abstract}





\section{Introduction}

The presence of small bodies in mean motion resonance is a ubiquitous feature in the Solar System. Most asteroids known to be in resonance have small to moderate inclination $\leq 40^\circ$. They can be part of large populations such as the asteroid belt with its Jupiter resonances or the Kuiper belt and its Neptune resonances. Temporary mean motion resonance capture can also occur such as with Centaurs in the outer planets' domain. More recently, it was found that mean motion resonance capture with Jupiter and Saturn occurs at large retrograde inclinations ($\geq 140^\circ$)  \citep{Morais&Namouni2013MNRAS}. One object is even orbiting  in the coorbital region of Jupiter with an inclination of $160^\circ$ \citep{Wiegert_etal2017Nature,NamouniMorais17c}. However, small bodies with polar orbits (i.e. with an inclination near $90^\circ$) have not been observed in mean motion resonance with a planet. This is most surprising as capture in polar resonances is known to have significant likelihood \citep{NamouniMorais15,NamouniMorais17}. In this Letter, we report that transneptunian object (471325) 2011 KT19, nicknamed Niku is currently in the 7:9 mean motion resonance with Neptune.

TNO Niku  (meaning rebellious in Chinese) was identified by \citet{Chen_etal2016ApJ} and has an orbital inclination of $110^\circ$. Another TNO, 2008 KV42 (nicknamed Drac),  with orbital inclination $103^\circ$ was identified by \citet{Gladman_etal2009ApJ}.  The origin of TNOs  with such high inclinations is a matter of debate.  They could  be part of a new reservoir in the Kuiper belt \citep{Gladman_etal2009ApJ} or possibly a sub-product of the gravitational sculpting on the extended scattered disk by an hypothetical planet \citep{Batygin&Brown2016ApJ}. The fact that these objects can be in mean motion resonances with the  planets may provide additional clues on possible formation mechanisms.

In this Letter, we start by reviewing briefly the work on the polar disturbing function \citep{Namouni&Morais2017MNRASb} that we need and then  apply it to Niku's orbit to identify the relevant argument of the 7:9 mean motion resonance with Neptune. We then generate a set of clones of Niku which are consistent with the observation's covariance matrix in order to analyse the stability and mean lifetime of  the resonant configuration. Finally, we assess the likelihood of other nearly polar TNOs being in a resonant configuration and discuss the implications of these results.

\section{The polar disturbing function and  Niku's resonance}
We have recently developed a disturbing function for nearly polar orbits to help us identify the possible resonances and dynamical behaviours at such inclinations \citep{Namouni&Morais2017MNRASb}. In general, the disturbing function is a series expansion of the gravitational interaction of two bodies that revolve around the Sun. The problem we faced for polar motion with the classical disturbing function for prograde motion  \citep{ssdbook}  and its companion for retrograde motion \citep{Morais&Namouni2013CMDA}, widely used in planetary dynamics,  is that their expansions are done with respect to nearly coplanar motion or equivalently for a relative inclination near zero. They are therefore not suitable for suggesting the relevant resonant arguments. When looking for resonances, we tend to favour the strongest. Then the classical disturbing function informs us that these are the pure eccentricity resonances of  argument $q \lambda-p \lambda^\prime+(p-q) \varpi $ where $\lambda$ and $\lambda^\prime$ are the small body  and planet's mean longitudes and $\varpi$ the small body's longitude of pericentre. The force amplitude of such resonances depend only on eccentricity and not inclination hence their strength. 

The polar disturbing function shows that for  a $p$:$q$ mean motion resonance, the possible arguments  are:
 \begin{equation} 
 \phi^{p:q}_k=q \lambda-p \lambda^\prime+(p-q) \Omega-k \omega \label{phi}
 \end{equation}
where $\Omega$ and $\omega$ are the small body's longitude of ascending node and argument of pericentre. The integer $k$ is even if $p-q$ is even or odd if $p-q$ is odd. The amplitude of resonant argument (\ref{phi})  is, at lowest order, proportional to $e^{|k|}$ and does not carry an inclination dependence despite the presence of the longitude of ascending node in the argument's expression.  Therefore, at lowest order in eccentricity, the 7:9 mean motion resonance has the possible arguments   $\phi^{7:9}_k=9 \lambda-7 \lambda' -2 \Omega-k \omega$ with $k=0$, $k=\pm2$, $k=\pm4$. 

In a preliminary result presented in \citet{Namouni&Morais2017MNRASb}   based on numerical integration of Niku's orbit including only the gravitational interaction of the Sun and Neptune moving on a circular orbit  we  remarked that Niku was likely to be captured  in 7:9 mean motion resonance with Neptune ($k=4$ argument).  

 In order to confirm this result we used  MERCURY \citep{Chambers1999} to  integrate the full equations of motion of Niku in 3 setups including: a) the 8 planets using the Burlisch-Stoer method with accuracy $10^{-12}$ over $\pm 200,000$ yrs ; b)  the 8 planets using the Hybrid method (a symplectic integrator that changes to the Bulirsch-Stoer method  to resolve close encounters at a distance that we chose as 10 planets' Hill's radii) and a time-step of 15 days over $\pm 50$ Myrs; c)  the 4 giant planets using the Hybrid method  with a time-step of 100 days over $\pm 500$ Myrs. The last integration period is of the order of Niku's half life \citep{Chen_etal2016ApJ}.  We find that in all simulations Niku is indeed  in the 7:9 mean motion resonance with Neptune  with resonant argument:
\begin{equation}
\phi^{7:9}_4=9 \lambda-7 \lambda' -2 \Omega-4 \omega \ .
\end{equation}

\begin{table*}
\centering
\caption{Sun-centred osculating elements of the nominal orbits  of multi-opposition TNOs with $110^\circ\ga I\ga70^\circ$, $a<100$~AU and $e<0.9$ at JD~$2457800.5$ with 1-$\sigma$ uncertainties both obtained from AstDys. The mean longitude is related to the mean anomaly $M$ by $\lambda=M+\omega+\Omega$.}
\label{table:1}
\begin{tabular}{|c|c|c|c|c|c|c|}
\hline \hline
& $a$ (AU) &  $e$  &  $I$ ($^\circ$)  &  $\Omega$ ($^\circ$) & $\omega$ ($^\circ$) & $M$ ($^\circ$)  \\
\hline
(471325) & $35.5895\pm 0.0016$ & $0.331647\pm 0.000029$ & $110.13335\pm 0.00012$ & $243.769738\pm 0.000029$ & $322.1991\pm 0.0049$ &  $30.4301\pm 0.0034$ 	\\
2008 KV42 & $41.385\pm 0.013$ & $0.49005\pm0.00022$ & $103.42415\pm0.00024$ & $260.91036\pm0.00029$ & $133.2999\pm0.0063$ &  $334.622\pm 	0.017$ 	 \\
2014 TZ33 & $38.3072\pm0.0041$ & $0.754497\pm0.000025$ & $86.005800\pm0.000078$ & $171.788143\pm0.000033$ & $159.02954\pm0.00084$ &  $2.67328\pm0.00043$ 	\\
2015 KZ120 & $46.0039\pm0.0053$ & $0.818012\pm0.000020$ & $85.556333\pm0.000054$ & $249.987343\pm0.000018$ & $66.26715\pm0.00081$ &  $357.64106\pm0.00044$  \\
(127546) & $67.612\pm0.022$ & $0.689297\pm0.000093 $ & $77.93897\pm0.00018$ & $90.364258\pm0.000046$ & $28.2126\pm0.0018$ &  $5.4153 \pm0.0025$ 	\\
2010 WG9 & $52.915\pm0.025$ & $0.64529\pm0.00015$ & $70.30809\pm0.00035$ & $92.06994\pm0.00025$ & $293.0287\pm0.0067$ &  $10.3002\pm 0.0082$ 	\\
\hline \hline
\end{tabular}
\end{table*}

\begin{figure*}
	\includegraphics[width=\textwidth]{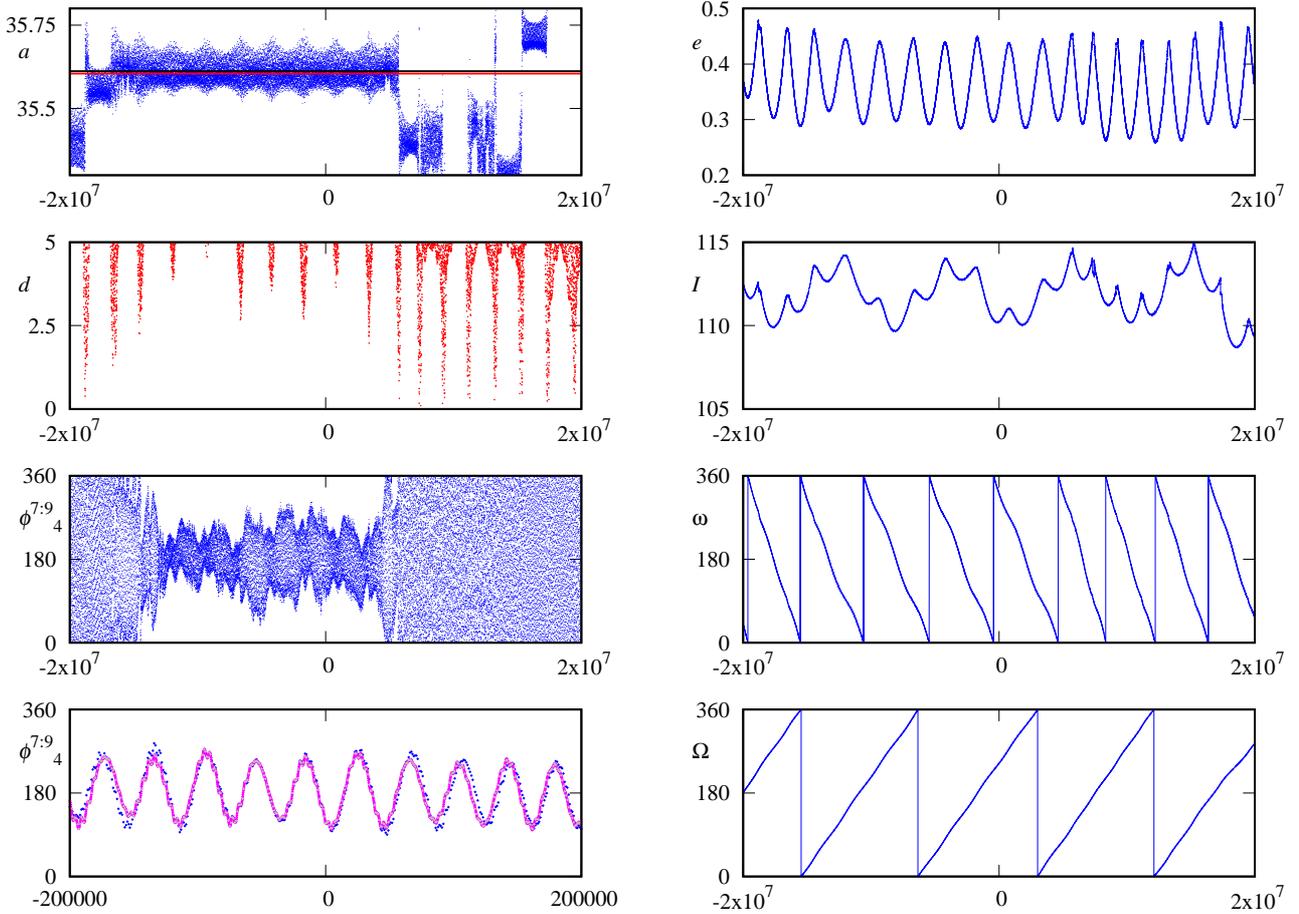}
    \caption{ Evolution of Niku's nominal orbit from Table ~\ref{table:1} in setup c). On the left panel: barycentric semimajor axis $a$ with location of the 7:9 resonance with Neptune (red horizontal line) and 1:18 resonance with Jupiter (black horizontal line); distance to Neptune $d$ (in units of the planet Hill's radii); resonant angle $\phi^{7:9}_4$ for $\pm 20$~Myrs  (3th panel) and for $\pm 200,000$~yrs (4th panel) in setup a) (grey), b) (magenta) and c) (blue). On the right panel: barycentric orbital elements $e$, $I$, $\omega$, $\Omega$.}
    \label{fig:1}
\end{figure*}

\begin{figure}
	\includegraphics[width=0.4\textwidth]{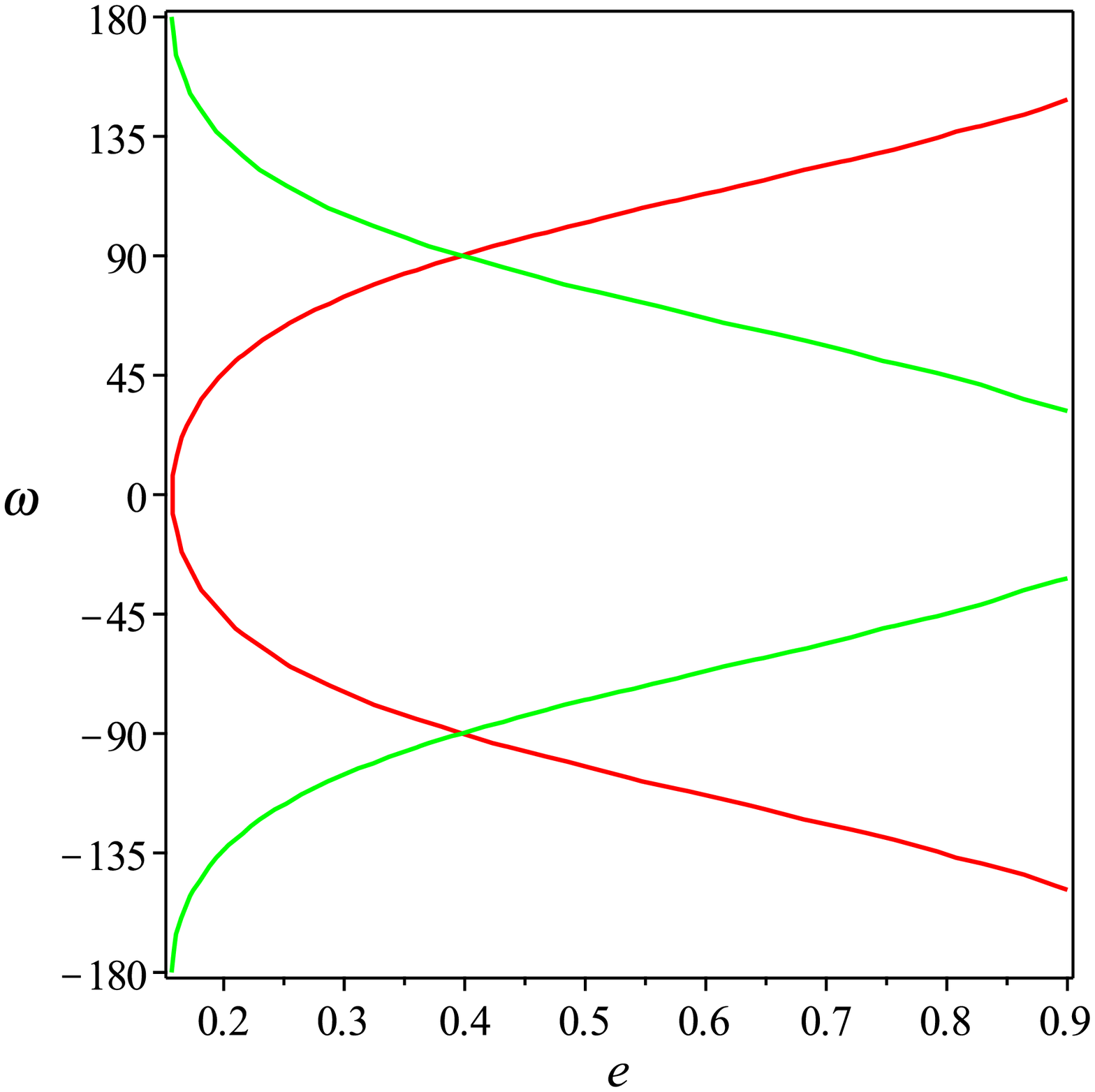}
    \caption{The $(e,\omega)$ values such that Niku's ascending node $r_+$ (red curve) or descending node $r_-$  (green curve) distances are at $30$~AU.  Niku's eccentricity values near present time, $0.5>e>0.3$, implies  values of $\omega\approx\pm 90^\circ$.}
    \label{fig:2}
\end{figure}

Figure~\ref{fig:1} shows the evolution of the barycentric orbital elements obtained by numerical integration of the mean (nominal) orbit  from AstDys\footnote{http://hamilton.dm.unipi.it/astdys} at epoch JD~$2457800.5$  (Table~\ref{table:1}).  The positions and velocities of the planets were obtained from JPL/HORIZONS\footnote{http://ssd.jpl.nasa.gov}. The blue curves show the evolution in setup c) while the grey and magenta curves  (4th left panel) show the evolution  of the resonant angle in setups a) and b), respectively. On  the $\pm200,000$~yrs timescale the evolution in setups a) and b) are undistinguishable but there are small differences in setup c) probably due to the secular frequencies being slightly different in  the simulations with the 4 or 8 planets.  Niku's barycentric semimajor axis is shown together with the location of nearby resonances with the giant planets (1st left panel).  The semi-major axis of the barycentric orbit has smaller amplitude oscillations than the Sun-centred one, thus it is more reliable in identifying the proximity to the web of nearby mean motion resonances \citep{Bannister_etal2016AJ}. The closest are the 7:9 resonance with Neptune (red horizontal line) and the 1:18 resonance with Jupiter (black horizontal line) which almost coincide. The $k=4$ argument of the 7:9 resonance with Neptune librates around $180^\circ$ with a period of about $40,000$ yrs (3rd and 4th left panels).   Resonance capture and exit occur due to nudges in semimajor axis caused by close encounters with Neptune (1st, 2nd and 3rd left panels) which may occur when Niku's distance to the Sun at the nodes, $r_{\pm}=a (1-e^2)/(1\pm e \cos{\omega})$, coincides with  Neptune's orbit.  Planetary close encounters are thus driven  by the secular period of the argument of pericentre, and due to the $180^\circ$ shift between the ascending and descending nodes, occur at 4 values: $\omega=90^\circ\pm\alpha$ and $\omega=-90^\circ\mp\alpha$.  Niku's current orbital parameters imply $\alpha\approx 0$ i.e. $\omega\approx\pm 90^\circ$  (Fig.~\ref{fig:2}) but  libration of $\phi^{7:9}_4$ around $180^\circ$ protects it from encounters with Neptune closer than 2 Hill's radii (2nd and 3th left panels).  Indeed, such close encounters may only occur when Niku is near the ascending or descending nodes ($\lambda\approx\Omega_{+}=\Omega$ or $\lambda\approx\Omega_{-}=\Omega+\pi$) and $r_{\pm}\approx 30$~AU (i.e.\ $\omega\approx\pm90^\circ$). Replacing these values into $\phi^{7:9}_4=180^{\circ}\pm\Delta\phi$ provides an estimate of the minimum angular distance between  Neptune and Niku while the resonance is maintained:
\begin{equation}
\Omega_{\pm}-\lambda' \approx \frac{  180^\circ -\Delta\phi}{7} \ .
\end{equation}
Hence, a resonance libration amplitude $2\,\Delta\phi<320^\circ$ prevents close encounters with Neptune at distances  less than 2 Hill's radii.\footnote{A similar procedure  is used to estimate the maximum libration amplitude of stable low inclination orbits in resonances but in that case the relevant measure is the angular distance  between the TNO's longitude of pericentre and Neptune's mean longitude, $\varpi-\lambda'$, where $\varpi=\omega+\Omega$ \citep{Malhotra1996AJ}.}

We also  integrated clones of Niku with the nominal orbit's parameters from Table~\ref{table:1} but semimajor axis deviated from the nominal value by $\pm 3.0\sigma_a$. This procedure has been proposed  by \citet{Gladman_etal2008}  as a check to confirm resonances in the Kuiper belt. Both   clones exhibit similar behaviour to the nominal orbit of Fig.~\ref{fig:1}.

Additionally, we performed numerical integrations  for $\pm 500$ Myrs in setup c) of a set of 100 clones  with orbital parameters generated   from a  multivariate normal distribution with mean equal to the  nominal  equinoctial  orbital elements' vector  and associated covariance matrix both  taken from AstDys at epoch  JD~$2457800.5$. All clones are currently in the 7:9 mean motion resonance with Neptune ($k=4$ argument) which further confirms that the resonance is robust. The resonance configuration lasts on average  $16\pm11$~Myrs  and, similarly to the nominal orbit's integration, the clones' capture and exit from the resonance occurs due to close encounters with Neptune.  Since Niku's perihelion (and minimum nodal distance) is currently   $23.5$~AU, encounters with Uranus occur at distances $>10$ Hill's radii and have negligible effect on the orbit.

\section{Are other nearly polar TNOs in resonance?}

Table~\ref{table:1} shows the nominal orbital elements of multi-opposition TNOs with $70^\circ\la I\la110^\circ$, $a<100$~AU and $e<0.9$ listed by the IAU Minor Planet Centre (MPC)\footnote{http://www.minorplanetcenter.net}. We briefly describe the behaviour  of the other TNOs in Table~\ref{table:1} with the purpose of assessing if they are in resonance. This study is not intended to be exhaustive.

The nominal barycentric semimajor axis of 2008 KV42  is near the 8:13 resonance with Neptune but there is no current libration. The resonant argument $k=9$ of the  $+3.0\sigma_a$  clone librates stably around $180^\circ$ but the $-3.0\sigma_a$ clone does not librate in the 8:13 resonance. This precludes confirmation of the resonance according to the criterion by \citet{Gladman_etal2008}. However, we observe that small nudges in semimajor axis due to close encounters with Uranus (the orbit's perihelion is currently $21.1$~AU) and Neptune at distances $\ga 3$ Hill's radii cause temporary libration ($10^5$ to $10^6$~yrs) in the 8:13 resonance ($k=9$ or $k=5$ arguments) for the nominal orbit and all  $\ga-1.5\sigma_a$ and $<+2.5\sigma_a$ clones. The  $\ga+2.5\sigma_a$ and $\la+3.0\sigma_a$  clones exhibit libration ($k=9$ argument) for the entire $10^7$~yrs numerical integration duration.   A reduction in the uncertainty of the orbit's semimajor axis which is  an order of magnitude larger than that of (471325) should help decide if 2008 KV42 is in resonance or not.  

The next two TNOs in Table~\ref{table:1} have the largest eccentricities in our sample and their nodes are currently  close to Saturn's orbit so they exhibit fast diffusion in semimajor axis due to close encounters with this planet.
2014 TZ33 nominal barycentric semimajor axis oscillates around the location of the 7:10 resonance with Neptune, 1:8 resonance with Saturn and 1:20 resonance with Jupiter. The associated resonant arguments exhibit intermittent behaviour  which excludes a stable resonance configuration.  The intermittency may be caused by interaction between the resonances as well as close encounters with Saturn. Such close  encounters eventually  cause ejection from the resonances' vicinity in a few thousand years.
2015 KZ120 has similar behaviour with close encounters with Saturn removing the orbit from the vicinity of the 9:17 resonance with Neptune, 4:15 resonance with Uranus and 1:26 resonance with Jupiter also in a few thousand years.

The semimajor axis and associated uncertainty for the remaining  TNOs in Table~\ref{table:1} is the largest in the sample.
The nominal orbit of 2010 WG9's librates intermittently in the 3:7 resonance with Neptune and 1:13 resonance  with Saturn. This could be due to interaction between the two resonances and close encounters at distances $\ga5$ Hill's radii of Uranus , while  close encounters with Neptune move the TNO's semimajor axis out of the present location. Planetary close encounters are driven by the argument of pericentre's precession timescale of around $10$~Myrs. The nominal orbit of (127546)  is closest to the 7:23  resonance with Neptune with brief intermittent  libration. Close encounters at distances $\ga 3$ Hill's radii with Uranus as well as close encounters with Neptune, both occurring on the argument of pericentre's precession timescale, cause the orbit's semimajor axis diffusion.  

We did not include in Table~\ref{table:1} a recently discovered nearly-polar TNO, 2017 CX33, as it has been observed for only one opposition. The nominal orbital parameters published on the IAU Minor Planet Centre imply that  it would  currently experience close encounters with Saturn near the orbit's descending node hence a stable resonance configuration would be unlikely.

\section{Conclusions}
 
We applied the recently developed disturbing function for polar orbits    \citep{Namouni&Morais2017MNRASb}   to show that TNO  (471325), also known as Niku, is currently in the 7:9 mean motion resonance with Neptune making it the first example of a Solar System object in polar resonance. The resonance configuration lasts on average $16\pm11$ million years and provides a temporary protection mechanism from close encounters with Neptune.  

We briefly analysed the possibility that  other nearly polar transneptunian objects may be in resonance with the planets. Of these, only 2008 KV42, also known as Drac, has a small probability ($\sim1\%$ with the present data) of being in the 8:13 resonance with Neptune. A better determination of the orbit's semimajor axis, whose current uncertainty is  an order of magnitude larger than that  of Niku, should help in deciding Drac's resonant status. 

Niku's nearly polar orbit at the 7:9 resonance  is  located between the 4:5 and 3:4 resonances with Neptune. The latter are both populated by objects with low inclination orbits ($I<20^\circ$) while the Plutinos in the nearby 2:3 resonance with Neptune have $I<40^\circ$ \citep{Gladman_etal2012AJ}. The presence of small objects on polar orbits with angular momenta lying near the Solar System's invariant plane is intriguing. Those in resonance have an increased protection from disruptive close encounters with the planets. Understanding their origin  will help constrain the processes of formation and evolution of the Solar System.

\section*{Acknowledgements}

This work has been funded by grant 2015/17962-5 of S\~ao Paulo Research Foundation (FAPESP).



\bibliographystyle{mnras}
\bibliography{Niku} 




\bsp	
\label{lastpage}
\end{document}